\documentclass[12pt]{article}
\usepackage{amsmath,amssymb,amsbsy,amsfonts,amsthm,latexsym,amsopn,amstext,
            amsxtra,euscript,amscd}
\title{{\large\bf Insecure primitive elements in an ElGamal signature protocol}}
\author{Omar Khadir\\
 Laboratory of Mathematics, Cryptography and Mechanics, Fstm\\
  University Hassan of Casablanca,   Morocco\\
e-mail: khadir@hotmail.com
}

\date{}
\begin{document}
\maketitle

 \baselineskip=20pt

\noindent\hrulefill

{\abstract \small Consider the classical ElGamal digital signature
scheme based on the modular relation  $\alpha^m\equiv y^r\, r^s\
[p]$. In this work, we prove that if we can compute a natural
integer $i$ such that $\alpha^i\ mod\ p$ is smooth and divides
$p-1$, then it is possible to sign any given document without
knowing the secret key. Therefore we extend and reinforce
Bleichenbacher's attack presented at Eurocrypt'96.

\noindent\hrulefill

 \vspace{0.3cm} \noindent {\small \bf Keywords :} \
{\small ElGamal signature scheme, public key cryptography,
cryptanalysis.}

\vspace{0.5cm}

\noindent {{\small \bf MSC 2010 : } {\small 94A60}

 \section{Introduction}

 \baselineskip=20pt
 It was in 1976 that Diffie and Hellman published their famous
paper "New directions in cryptography" [4]. For the first time in
communication history, they provided us with a mechanism that
guarantees the confidentiality of documents and data we like to
exchange over a public and insecure channel. This event is at the
origin of the public key cryptography [4,14,13]. Since then, many
original cryptographical methods were conceived and proposed to
solve a variety of communication problems like identification,
authentication, integrity or 0-knowledge proof. However, the most
important field in public key cryptography is probably the digital
signature protocol. Its requirement in e-business for funds
transferring, makes it a sensitive question. Let us recall the
principle. For the user Alice we prepare two kind of keys. The
first, $y$, is public and must be largely diffused to the other
users. The second, $x$, is private and must be kept secret. When
Alice decides to sign a document $M$, she has to solve a difficult
problem, in general a mathematical equation. This problem is
depending of  Alice public key $y$ and of the document $M$. It is
constructed in a way such that nobody, except Alice,  can solve
it. With the help of her secret key $x$, Alice is able to give the answer.\\
The equation is based on a hard question  in mathematics like
factorization or discrete logarithm  problem. We cannot forge
Alice signature, but anyone like a judge can verify that the
solution  she gives is valid.

Let $p$ be a prime number and $\alpha$ a primitive element modulo
$p$. The discrete logarithm problem consists of solving the
modular equation $\alpha^x\equiv \beta \ [p]$, where $\beta$ is a
fixed  integer and $x$ is the unknown variable. In 1978, Pohlig
and Hellman [12] elaborated an efficient algorithm when $p-1$ is
$B-$smooth. In 1985, ElGamal [6] proposed a public key
cryptosystem and one of the first digital signature protocols both
based on the discrete logarithm. Nobody knows how he found his
difficult signature equation. Several variants of the signature
scheme were developed [15, 5, 10 table 11.5 p.457,7,9]. In 1996,
Bleichenbacher [2,3] built an attack that relies on Pohlig and
Hellman algorithm if ElGamal signature parameters are not properly
chosen. In 1999, Kuwakado and Tanaka [9] proved that, when we use
ElGamal method to sign two documents, if the secret nonces $k_1,
k_2$ are less than the square root of the prime modulus $p$, then
we can compute the secret key of the signer and break all the
system. In 2011, the author slightly extended Bleichenbacher's
attack [8].

Let $\alpha^m\equiv y^r\, r^s\ [p]$ be the ElGamal classical
signature equation. In this work, we show that if we can compute a
natural integer $i$ such that $\alpha^i\ mod\ p$ is $B-$smooth and
divides $p-1$, then it is possible to sign any given document
without knowing the secret key. As a consequence, we prove that if
$(p,\alpha,y)$ is Alice public key, and if one the four positive
integers $\alpha$, $p-\alpha$, $\displaystyle \frac{1}{\alpha}\
mod\ p$ or $\displaystyle -\frac{1}{\alpha}\ mod\ p$ is B-smooth
and divides $p-1$, then it is possible to sign any message without
knowing Alice private key. Therefore we extend and reinforce
Bleichenbacher's attack
presented at Eurocrypt'96.\\
 Note also, that our work tends to
confirm, what was mentioned by many authors: it is certainly
easier to break ElGamal signature scheme than to solve the
discrete logarithm problem.

Our paper is organized as follows. In section 2 we briefly recall
the classical ElGamal signature scheme. Section 3 is devoted to
the review of Bleichenbacher's attack [2,3]. Our contribution is
presented in section 4. We conclude in section 5.

Throughout this article, we will adopt  ElGamal paper notations
[5]. $\mathbb{Z}$, $\mathbb{N}$ are respectively the sets of
integers and non-negative integers. For every positive integer
$n$, we denote by $\mathbb{Z}_n$ the finite ring of modular
integers and by $\mathbb{Z}_n^*$ the multiplicative group of its
invertible elements. Let $a,b,c$  be three integers. The great
common divisor of $a$ and $b$ is denoted by $gcd(a,b)$. Two
numbers $a$ and $b$ are said to be coprime if $gcd(a,b)=1$. We
write $a\equiv b$ $[c]$ \ if $c$ divides the difference $a-b$, and
$a=b\ mod\ c$ if $a$ is the remainder in the division of $b$ by
$c$. The positive integer $a$ is said to be B-smooth [10, p.92],
$B\in\mathbb{N}$, if every prime factor of $a$ is less than or
equal to the bound $B$. Generally, parameter $B$
depends of the computer power.\\

\section{Classical ElGamal signature}

 \baselineskip=20pt

In this section we recall the basic ElGamal signature scheme [6, 16 p.287, 10 p.454, 11 p.183].\\
 1. Alice chooses three numbers:

\hspace{1cm}- $p$, a large prime integer.

\hspace{1cm}- $\alpha$, a primitive element (or a generator) [10,
p.69] of the finite multiplicative group $\mathbb{Z}^{*}_{p}$

\hspace{1cm}- $x$, a random element belonging to the set  $\{2,3,...,p-2\}$.\\
Then she computes $y = \alpha^{x}$ $mod$ $p$. Alice public keys are $(p,\alpha,y)$, and $x$ is her private key.\\
 2. To sign the message m, Alice needs to solve the equation :
$$ \alpha^m\equiv  y^r r^s\ [p]\eqno{\bf (1)}$$
where $r, s$ are the unknown variables.\\
Alice fixes arbitrary $r$ to be $r= \alpha^{k}$ $mod$ $p$, where
$k$ is chosen randomly and invertible modulo $p-1$. Equation (1)
is then equivalent to :
$$ m \equiv xr + ks\  [p-1]          \eqno{\bf(2)}$$
As Alice knows the secret key $x$, and as the integer $k$ is
invertible modulo $p-1$, she computes the second unknown variable
$s$:  $\displaystyle s \equiv\frac{m-xr}{k}\ [p-1]$\\
3. Bob can verify the signature by checking that
 congruence (1) is valid for the variables $r$ and $s$ given by Alice.

\vspace{0.2cm}\noindent  To avoid some attacks, instead of signing
a message $M$, it is more secure to apply a hash function $h$,
like SHA1 [16 p.137, 10 p. 348], and compute $m=h(M)$ before
signing the hashed value $m$.

 \baselineskip=20pt

\section{Bleichenbacher's attack}
In this part we recall Bleichenbacher's remarquable attack
presented at the Eurocrypt'96 conference [2]. Here, of course, we
use the corrected version [3].

Let $(p,g,y_A)$ be Alice public key in an ElGamal signature
scheme, and $x_A$ his private key.

\noindent {\bf Theorem 1. [3]} let $p-1=bw$ where $b$ is smooth
and let $y_A\equiv g^{x_A}\ (mod\ p)$ be the public key of user A.
If $r$ and $k$ are known such that $r\equiv \alpha^k\equiv cw\
(mod\ p)$ with $0<c<b$ then it is possible to generate a valide
ElGamal signature $(r,s)$ for all $h$ with $h\equiv x_Ar\ (mod\
gcd(k,p-1))$ can be found. In particular when $r$ is a generator
of $\mathbb F_p^*$ then it is possible to generate an ElGamal
signature for all $h$.

\vspace{0.2cm} \noindent Theorem 1 has an immediate practical
consequence :

\vspace{0.1cm} \noindent {\bf Corollary 1. ([3])} If $\alpha$ is
$B-$smooth and divides $p-1$ then it is possible to generate a
valid ElGamal signature on an arbitrary value $h$ if $p\equiv 1\
[4]$ and on one half of the values $0\leq h<p$ if $p\equiv 3\
[4]$.

\vspace{0.2cm} \noindent when $p\equiv 1\ [p]$, we easily derive
the following algorithm and we will exploit it in an illustrative
example of our own attack.

\vspace{0.1cm} \noindent {\bf Algorithm 1.}

\noindent {\bf 1-} Input $(p,\alpha,y)$; \{$\alpha$ is $B-$smooth
and divides $p-1$, $p\equiv 1\ [4]$\}\\
{\bf 2-} Input $m$;  \{$m=h(M)$ where $M$ is the message to be signed.\}\\
 {\bf 3-} $k\longleftarrow (p-3)/2$;\\
{\bf 4-} $r\longleftarrow\alpha^k\  mod\  p$; \{ $r$ is is the
first parameter of the digital signature. We also have
$r:=(p-1)/\alpha$. \}\\
{\bf 5-} $w\longleftarrow(p-1)/\alpha$; \\
 {\bf 6-} $b\longleftarrow \alpha^w\ mod \ p$; \{$b$ is a generator of a suitable subgroup $H$\}.\\
{\bf 7-} $B\longleftarrow y^w\ mod \ p$; \{$B$ is an other element of $H$\}.\\
 {\bf 8-} $x_0\longleftarrow x$; \{ $x$ is a solution to the easy discrete logarithm problem  $b^x\equiv B\ [p]$, since
 the Pohlig and Hellman algorithm [12] is efficient. \}\\
{\bf 9-} $\displaystyle s\longleftarrow\frac{h(M)-rx_0}{k}\ [p-1]$; \{$s$ is the second parameter of the digital signature. \}\\
{\bf 10-} Output $(r,s)$. \{ The couple $(r,s)$ is the ElGamal
digital signature without using Alice private key $x$.\}

\vspace{0.2cm} \noindent In 2011, Corollary 1 was extended by the
author to the next more general result:

\vspace{0.1cm}
 \noindent {\bf Theorem 2. [8]}
Let $(p,\alpha,y)$ be Alice public key in an ElGamal signature
protocol. An adversary can forge Alice signature for any
given message if one of the following conditions is satisfied :\\
a) $p\equiv 1\ [4]$, $\alpha$ is B-smooth and divides $p-1$. \\
b) $p\equiv 1\ [4]$, $\displaystyle \frac{1}{\alpha}\ mod\ p$ is B-smooth and divides $p-1$. \\
c) $\alpha^2$ is B-smooth and divides $p-1$.

 \section{Our contribution}

We start this section by describing our main result which is a
significant extension of Bleichenbacher's Corollary 1. Throughout
this part, for more clarity and without loss of generality, we
always suppose that the prime modulus $p$ in equivalent to 1
modulo 4. When $p\equiv 3\ [4]$ all our results still valid but
only for documents $M$ such that the integer $m=h(M)$ has a fixed
parity.

\vspace{0.2cm}
 \noindent
{\bf Theorem 3.} Let $(p,\alpha,y)$ be Alice public key in an
ElGamal signature protocol. Suppose that $p\equiv 1\ [4]$. If we
can compute a natural integer $i$, coprime to $p-1$, such that
$\alpha^i\ mod\ p$ is $B-$smooth and divides $p-1$, then it is
possible to generate a digitale signature for any given document
without knowing Alice private key. \proof Let $M$ be  the message
that we would like to sign and $m=h(M)$ be its hashed value. We
must find two unknown integers $r$ and $s$ such that
$\alpha^m\equiv y^r\,r^s\ [p]$.\\
Let $i$ be a natural integer coprime to $p-1$ such that $\alpha^i\
mod\ p$ is $B-$smooth and divides $p-1$. ElGamal digital signature
Equation (1) is equivalent to
$$\alpha^{im}\equiv y^{ir}\,r^{is}\ [p] \eqno {(3)}$$
If we set  $\beta=\alpha^i\ mod\ p$, $z=y^i\ mod\ p$, $u=r$ and
$v=is\ mod\ (p-1)$, we obtain the new modular equation
$$\beta^m\equiv z^u\,u^v\ [p]\eqno {(4)} $$
Since $gcd(i,p-1)=1$, the element $\beta^i\ mod\ p$ is a primitive
root. As $\beta^x\ mod\ p=z$, where $x$ is Alice secret key, the
triplet $(p,\beta,z)$ can be seen as the public key of an
imaginary user in an ElGamal signature protocol. We do not need
the private key $x$. For any given document $M$, by Corollary 2,
it is possible to solve equation (4) and to find the unknown
variables $u$ and $v$. Therefore, we generate a signature by
giving the couple  $r=u$ and $\displaystyle s=\frac{v}{i}\ mod\
(p-1)$.

 \hspace{14.5cm} $\qed$

 \vspace{0.2cm} \noindent Observe that a trapdoor could be hidden in the generator $\alpha$
by choosing it such that $\alpha^i\ mod\ p$ is B-smooth and
divides $p-1$, with a large exponent $i$.\\
 To illustrate our technique, let us give a numerical
example.

 \vspace{0.3cm} \noindent {\bf Example 1.} Assume that $p=1597$,
 $\alpha=11$ and $y=159$. The secret key  $x=856$ is ignored. \\
 Suppose that we want to sign the message $M$ such that $m=h(M)=1234$, where $h$ is a hash function like SHA1.
 Observe, first, that Bleichenbacher's attack cannot be
 mounted here. On another side, conditions a) and b) in Theorem 2 are not satisfied. \\
Let us therefore apply our method. With the help of a computer, we
find that the smallest positive exponent $i$ such that
$\beta=\alpha^i\ mod\ p$ divides $(p-1)$ is $i=275$. As $z=y^i\
mod\ p=1287$, we determine the public key of a fictive user
$(p,\beta,z)=(1597,38,1287)$. Obviously $\beta$ is $B-$smooth.
Algorithm 1 gives us the signature $(u,v)=(42,1202)$. As $u=r$ and
$v= is \ mod\ (p-1)$, we obtain $(r,s)=(42, 370)$. So, we have
signed the message $M$ such that $h(M)=1234$ without using Alice
private key $x$. Any verifier can check that the ElGamal  modular
equation (1) is valid.

 \vspace{0.3cm} \noindent Assume that $(p,\alpha,y)$ is Alice public
 key. In somehow, our result in Theorem 3 means that, to break the ElGamal digital signature system,
  it is not needed to have $p-1$ a multiple of $\alpha$ as it is claimed  by Bleichenbacher [2,3],
   but it suffices to have  $p-1$ a multiple of anyone of the primitive elements modulo $p$.
   Next corollary  is another extension.

\vspace{0.2cm} \noindent {\bf Corollary 2.} Let $(p,\alpha,y)$ be
Alice public key in an ElGamal signature protocol. Suppose that
$p\equiv 1\ [4]$. If one among the four positive numbers $\alpha$,
$p-\alpha$, $\displaystyle \frac{1}{\alpha}\ mod\ p$ or
$\displaystyle \frac{-1}{\alpha}\ mod\ p$, is $B-$smooth and
divides $p-1$, then it is possible to generate a signature for any
given document without knowing Alice private key. \proof For
$\alpha$ and $\displaystyle \frac{1}{\alpha}\ mod\ p$  apply
respectively Corollary 1 and Theorem 2. Let us study the two cases
corresponding to $p-\alpha$ and $\displaystyle -\frac{1}{\alpha}\
mod\ p$. The even integer  $p-1$ can be decomposed as
$p-1=2^k\,l$, where $k,\, l$ are the two easily computable natural
numbers such that $k\geq 2$ and $l$ is odd.  Fermat little theorem
gives the modular relation $\alpha^{p-1}\equiv 1\ [p]$. As the
order of the primitive element $\alpha$ is $p-1$, looking at the
factorization of $\alpha^{p-1}-1$ modulo $p$, we necessary have
$\alpha^{2^{k-1}l}\equiv -1 \ [p]$ which implies
$\alpha^{2^{k-1}l+1}\equiv -\alpha \ [p]$ and $\displaystyle
\alpha^{2^{k-1}l-1}\equiv -\frac{1}{\alpha}  \ [p]$.  Since
$gcd(\alpha^{2^{k-1}\,l\pm 1},p-1)=1$, the proof is achieved by
immediate application of our theorem 3.

 \hspace{14.5cm} $\qed$

 \vspace{0.3cm} \noindent There is a well known particular situation for the generator choice:
  "Choosing $\alpha=2$ is exceptionally bad" [2,3,1,10 p.456]. We extend the case:

\vspace{0.2cm} \noindent {\bf Corollary 3.} Let $(p,\alpha,y)$ be
Alice public key in an ElGamal signature protocol. Suppose that
$p\equiv 1\ [4]$. It is possible to forge Alice digital signature
for any given message $M$ if we have one of the two conditions:\\
i) $\alpha=2$.\\
 ii) Number $2$ is a primitive element of
the multiplicative group $\mathbb{Z}_p^*$ and the positive
exponent $i$ such that $\alpha^i\equiv 2\ [p]$ is computable.
 \proof Similar to the justification of Theorem 3.

 \hspace{14.5cm} $\qed$

\section{Conclusion}
In this paper, we determined new primitive elements of the
multiplicative finite group $\mathbb Z_p^*$, $p$ prime, for which
ElGamal digital signature scheme is no more secure. We therefore
made an extension of the old and remarquable result presented by
Bleichenbacher at Eurocrypt'96.


\end{document}